\documentstyle[12pt]{article}
\begin{document}
\title{Magnetic Oscillations and Maxwell Theory}
\author{Artur Sowa\\
        Department of Mathematics\\
        Graduate Center of The City University of New York\\
        33 West 42nd Street\\
        New York, NY 10036\\
         sow@clsc2.gc.cuny.edu}
\date{ }
\maketitle
\newtheorem{tw}{Main Inequality}
\renewcommand{\l}{\bigtriangleup_{A}}

\begin{abstract}\small
We explore the possibility of using the Kaluza-Klein geometry of Riemannian
Submersions to modify the classical Maxwell Theory.
We further argue that the resulting modification of Electromagnetism---cf.
equations (\ref{main}) as well as (\ref{syst1}) and (\ref{syst2})---may 
be interesting in the context of, among other topics, magnetic oscillations
in metals.

\end{abstract}
\begin{enumerate}
\item 
The Maxwell equations are geometric and can be expressed as
equations  defining a principal $U(1)$ connection. Such a connection $A$
is a differential $1$-form on the total space of a $U(1)$-bundle, which
in addition satisfies
\begin{equation}
          R_{\theta}^* A=0 ,\qquad
          A(\frac{\partial}{\partial\theta})=1.
\label{con}
\end{equation}
Here $\theta\in U(1)$, $R_{\theta}$ denotes the $U(1)$-action on the
bundle
and $\frac{\partial}{\partial\theta}$ is the canonical vertical vector
field induced by this action.
Now if $F_{A}=dA$ is the curvature, then (in the absence of nonstatic
charge distributions) the Maxwell equations become
\begin{equation}
             dF_{A}=0,\qquad
             d^*  (f F_{A})=0,
\end{equation}
where $f$ is the so called material constant (like the dielectric
constant, the inverse of the magnetic permeability, etc.)
\footnote{In what follows all physical constants irrelevant for our 
discussion are set to $1$.}

\item It has been established by the Aharonov-Bohm experiment that the
connection itself has a physical meaning.
Thus we see that the equations (\ref{con})
are themselves a part of Maxwell Theory. Clearly, the first of the equations
states $U(1)$-symmetry. What is the significance, if any, of
$A(\frac{\partial}{\partial\theta})=1$, though?---We will try to suggest 
an answer to this question.

Let us recall that  principal connections  also have another
interpretation. A principal bundle naturally distinguishes the vertical
direction $\frac{\partial}{\partial\theta}$, but distinguishing a
complimentary distribution of horizontal planes requires an additional
bit of structure --- a connection. Such a horizontal distribution is
usually described as $\ker A$.

If we replace $A$ by $\tilde{A}=f\cdot A$, where $f$ is a positive
function constant along the 
fibers, then $\ker \tilde{A}=\ker A$, and
$\tilde{A}$ defines the same horizontal distribution as $A$.
We will call $\tilde A$ a connection with amplitude.

It has to be decided, which is the physical notion of a connection?
Is it the ``pure'' horizontal plane distribution (and the Aharonov-Bohm
experiments do not go beyond it), or is it the $1$-form $A$ or maybe
$\tilde{A}$ ---  i.e. there are {\em a priori} more than just one choice of 
$A$?

We argue in this article that the choice of $\tilde{A} $ is a very
interesting one. Moreover, we propose to interpret $f=\tilde{
A}(\frac{\partial}{\partial\theta})$ as a 
material ``constant", which in addition  will be allowed to vary 
according to a certain rule.

\item In the meantime, we want to ask if there is a natural candidate
for a system of geometric equations  for $fA$, such that the Maxwell
equations are a ``limiting'' case.

Kaluza and Klein were the first to observe that the structure induced by
a connection on a principal fiber bundle allows us to lift the metric
tensor of space-time to a metric tensor on the total space of the
bundle.

The tangent
space of the total space of a principal bundle $\pi :P\rightarrow M$ at
each point decomposes into the vertical, i.e. tangent to the
fiber, and the horizontal part defined by the connection. We declare
them orthogonal. The scalar product $\mu=\mu (A)$ is defined by
\[ \mu\left(X,Y\right) = A(X)A(Y) +
                         g\left(\pi _{\star}X,\pi _{\star}Y\right), \]
where $g$ is the metric tensor on $M$.
 
This metric defines the Laplace-Beltrami operator on forms
\[ \l = d\delta +\delta d ,\]
where $d$ is the exterior derivative and its formal adjoint $\delta$
acting on forms of
degree $q$ is defined by $\delta = (-1)^{dimP(q+1)+1}\star d\star$
 (where $\star$ denotes the $\mu$-Hodge star and
therefore depends on $A$).
 
We postulate that the elliptic eigenvalue problem
\begin{equation}
\l (fA)=\nu fA
\label{main}
\end{equation}
is a natural choice for a description of static states of the
connection with amplitude.
This equation was introduced in \cite{sow1} and further studied in
\cite{sow2} (with a wrong numerical constant) and \cite{sow3}.

\item Our aim now is to explicitly evaluate this Laplacian on the form
$fA$.  We assume for simplicity that $M$ is orientable,
but this
assumption is not necessary. Also $\dim M=m$.
 
Let us first notice that $dV_{P} = dV_{M}\wedge A$, where
$dV_{P}$ is the volume element of $P$ and $dV_{M}$ is the pull-back
of the volume element of $M$ to $P$. (In what follows we will often
identify differential forms on $M$ with their pull-backs to $P$.)
 Thus, in particular, $dV_{P}$ does not depend on the choice
of a connection $A$. Indeed, if $A'$ is another connection, then
$A'=A+da$ and $dV_{M}\wedge A'=dV_{M}\wedge A +dV_{M}\wedge a =
dV_{M}\wedge A$.
Now, if $\alpha$ is a horizontal $p$-form on $P$, i.e. $\alpha
(\frac{\partial}{\partial\theta})=0$, then we have
\[ \star\alpha = \star_{M}\alpha\wedge A,
\]
\[ \star (\alpha\wedge A) = (-1)^{m-p}\star_{M}\alpha ,
\]
where $\star_{M}$ is the Hodge star of $g$ (on $M$).
In particular $\star (fA)= (-1)^{m}fdV_{M}$ and, since $f$ is constant along
fibers, we see that the $1$-form $fA$ is co-closed.
 We have
 
\[ d(fA)=df\wedge A +fF_{A},
\]
\[  \star d(fA)= (-1)^{m-1}\star _{M}df + f\star _{M}F_{A}\wedge A,   
\]
\[  d\star d(fA)= (-1)^{m-1}d\star _{M}df + d(f\star _{M}F_{A})\wedge A
+(-1)^{m-2}f\star _{M}F_{A}\wedge F_{A}, \]    
\[\delta d (fA)=(-1)^{m-1} \star d\star d(fA) =
(-\bigtriangleup f +|F|^2 f)\wedge A + (-1)^{m}\delta _{M}(fF_{A}). \]
 
Thus the equation (\ref{main}) is equivalent to the following system of
equations (on M)
 
\begin{equation}
 \delta _{M} (fF_{A})=0
\label{syst1}   
\end{equation}
 
\begin{equation}
 -\bigtriangleup f +|F_{A}|^{2}f=\nu f.
\label{syst2}
\end{equation}
(Note that in this paper $\bigtriangleup \leq 0$ on functions, and
$\bigtriangleup \geq 0$ on $1$-forms.)
 
It should be noted that these equations are {\em not} the Euler-Lagrange
equations corresponding to  a functional in the two variables $f$ and $A$.

\item It is clear that $f$ plays the role of a ``material constant".
It has long been known from experiments that at low temperatures and
high magnetic fields, material constants lose their meaning as 
simple coefficients of proportionality. It was determined that in such
conditions the relation between, say, magnetic field $H$ and magnetic
induction $H= \frac{1}{\mu} B$ does no longer hold and should be
replaced by a more general relation $H=H(\mu ,B)$. (This one too, though, 
has a 
very limited range of applications.)  Moreover, in such extreme conditions
new phenomena have been observed, which could not be explained by the
Maxwell Theory and were attributed quantum character. Here we have in
mind all kinds of magnetic oscillations, e.g. de Haas-van Alphen or
Shubnikov-de Haas oscillations (see \cite{sch}), as well as the so called 
Quantum Hall
Effect or Fractional Hall Effect (see \cite{pr}). We will argue in the 
remaining 
parts of this article, that oscillatory phenomena are naturally built in
the system (\ref{syst1}), (\ref{syst2}). It should also be mentioned
that this system of equations, just as any other gauge theoretic
system of equations, contains information about the topological quantum
numbers. These could be later applied to modeling the QHE.

\item We want to understand basic features of solutions of the system
(\ref{syst1}), (\ref{syst2}) in the case of two dimensional
noncompact manifolds.
In this case $\delta (fF)=0$ implies $\star F=\frac{B}{f}$ for
a constant $B$. Also, in this case $dF=0$ and $F$ is a
curvature of a certain connection $A$.
Thus our system of equations reduces to
\begin{equation}
 -\bigtriangleup f +\frac{B^2}{f}=\nu f.
\label{twodim}
\end{equation}

Let us emphasize the fact that $\bigtriangleup f$ depends on the choice of
geometry, more strictly---the Riemannian metric tensor $(g_{ij})$.

We now show results of computer simulations for different choices
of geometry. All figures present solutions $f$ of the equation 
(\ref{twodim}) which are assumed to be
radially symmetric, or to depend on one spatial variable only, and satisfy
$f(0)=1, f'(0)=0$. The parameters $B$ and $\nu$ are set to convenient 
values, but do not affect the qualitative properties of the pictures too
radically.
We point out that the shape of the derivative of $f$ in all cases resembles
the distorted sinusoidal pattern. This pattern is characteristic for magnetic
interactions of external fields with ferromagnetics. One of the figures 
shows how the distortion depends on $\nu$. 

 The last figures show dependence of radially symmetric solutions 
with $f(0)=1, f'(0)=0$ on $\nu ^{\frac{1}{2}}$, i.e. we plot f(0.5) 
against $\nu ^{\frac{1}{2}}$ (fixed $B=1$, relative scale, undersampled 
data).

Let us also mention here that simulated
solutions of the system (\ref{syst1}), (\ref{syst2}) on manifolds of
three
and four dimensions (with special Ansatzen like e.g. the t'Hooft Ansatz)
display similar qualitative properties.

%%%%%%%%%%%%%%%%%%%%%%%%%%%%%%%%%%%%%%%%%%%%
%%%%%%%%%%%%%%%%%%%%%%%%%%%%%%%%%%%%%%%%%%%%
%%%%%%%%%%%%%%%%%%%%%%%%%%%%%%%%%%%%%%%%%%%%
%%%%%%%%%%%%      PICTURES  %%%%%%%%%%%%%%%%
%%%%%%%%%%%%%%%%%%%%%%%%%%%%%%%%%%%%%%%%%%%%
%%%%%%%%%%%%%%%%%%%%%%%%%%%%%%%%%%%%%%%%%%%%
%%%%%%%%%%%%%%%%%%%%%%%%%%%%%%%%%%%%%%%%%%%%

\vspace{1 cm} \vbox spread 3in{}
 \includegraphics{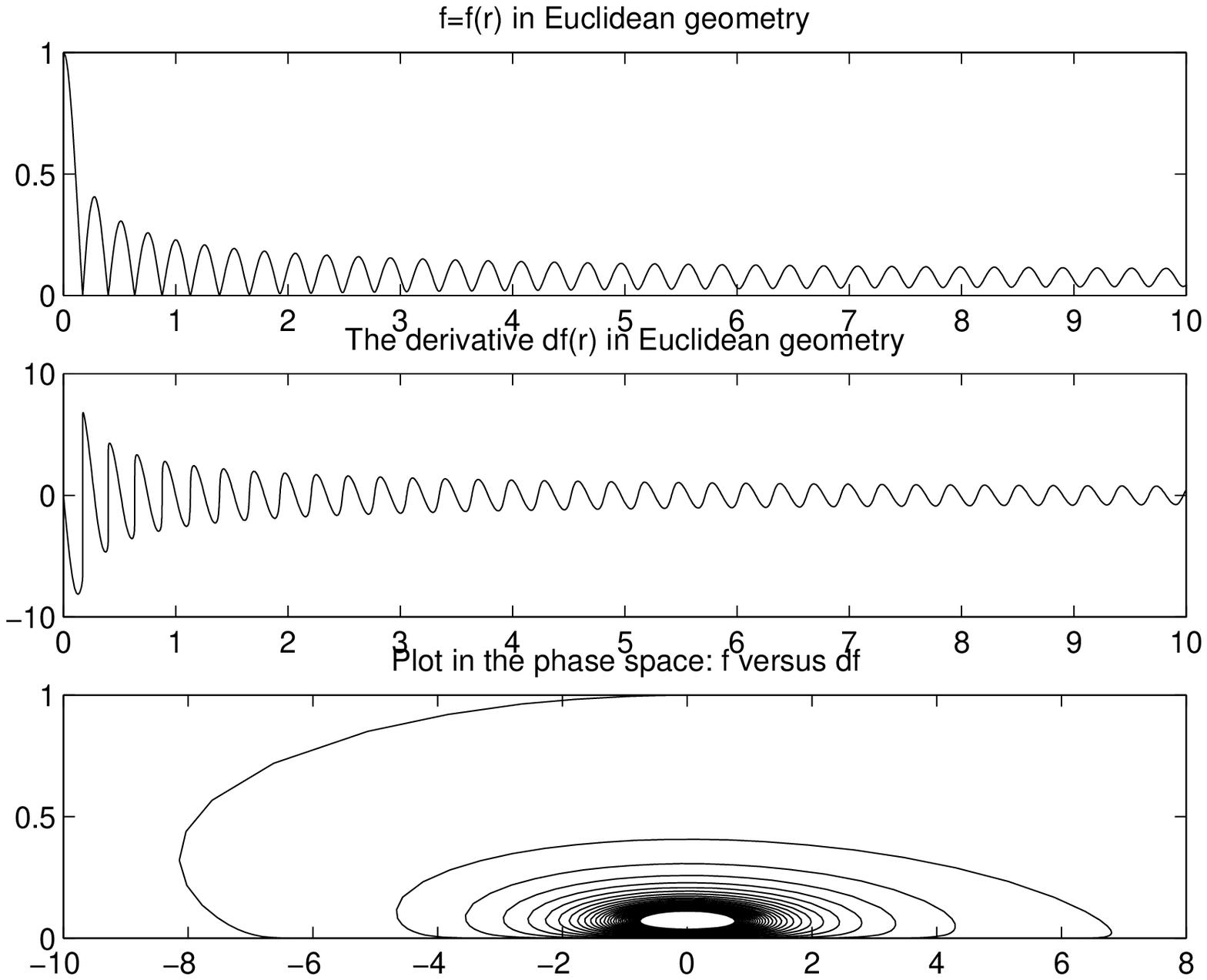}
 
\vspace{1 cm} \vbox spread 3in{}
 \includegraphics{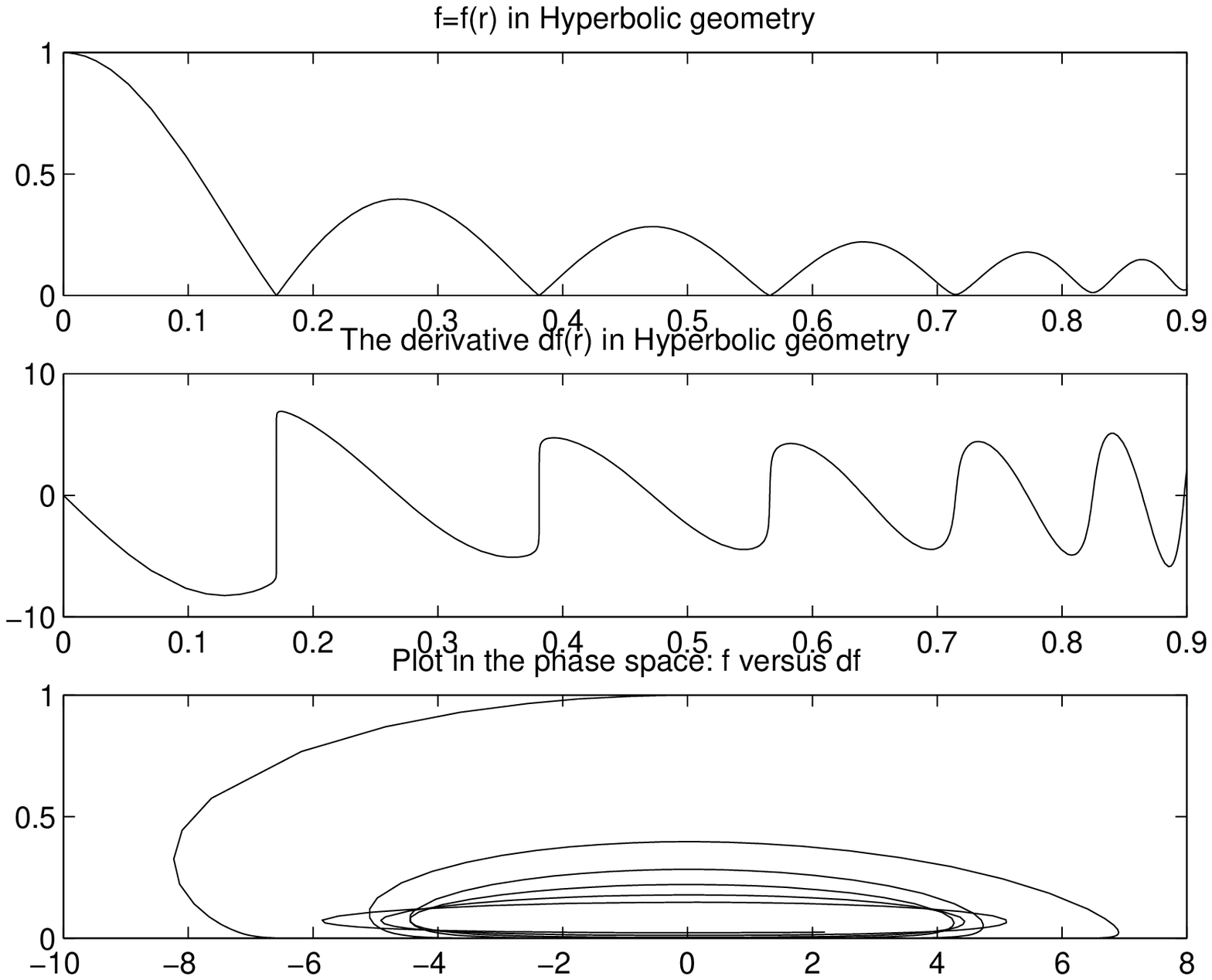}

\vspace{1 cm} \vbox spread 3in{}
 \includegraphics{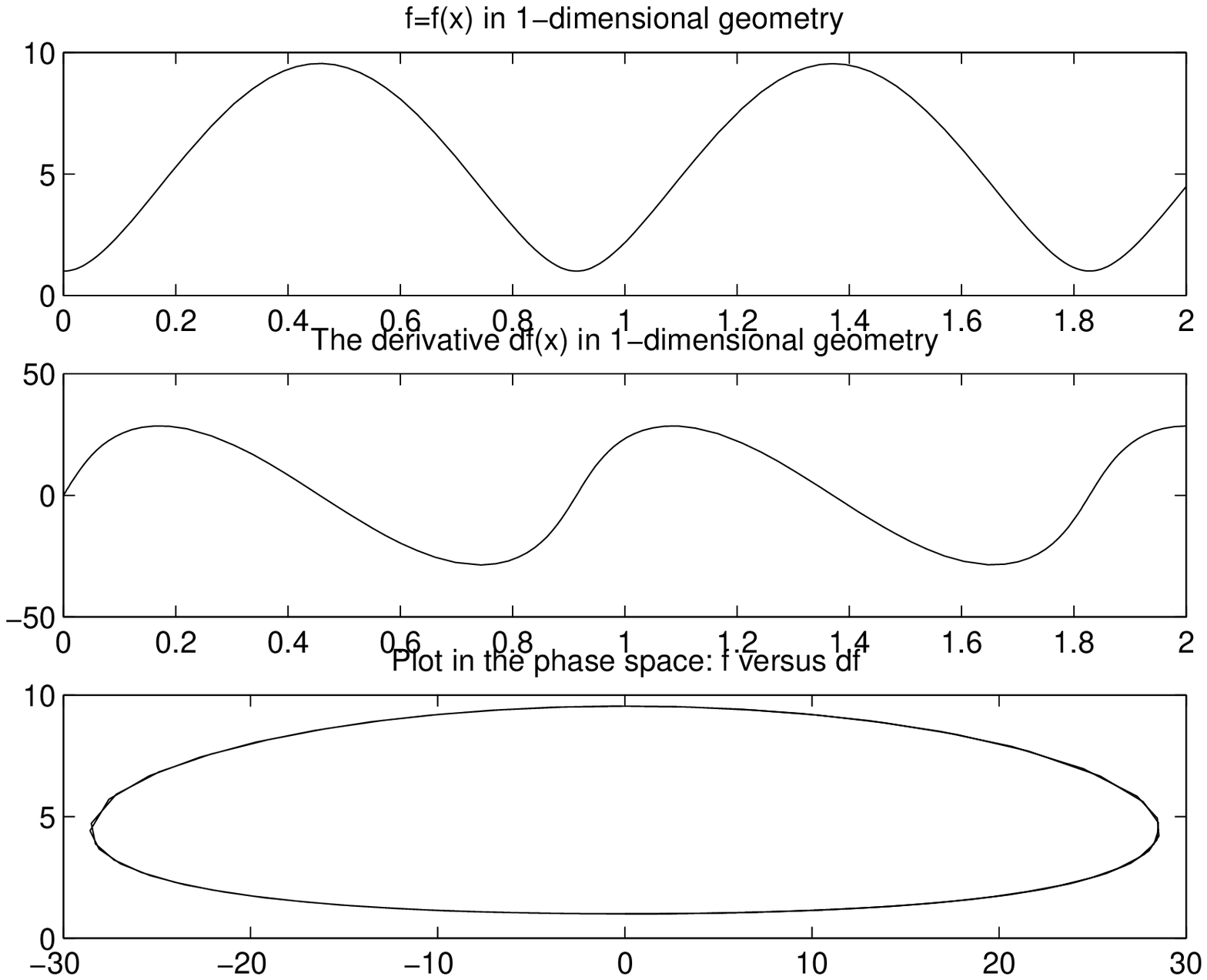}
 
\vspace{1 cm} \vbox spread 3in{}
 \includegraphics{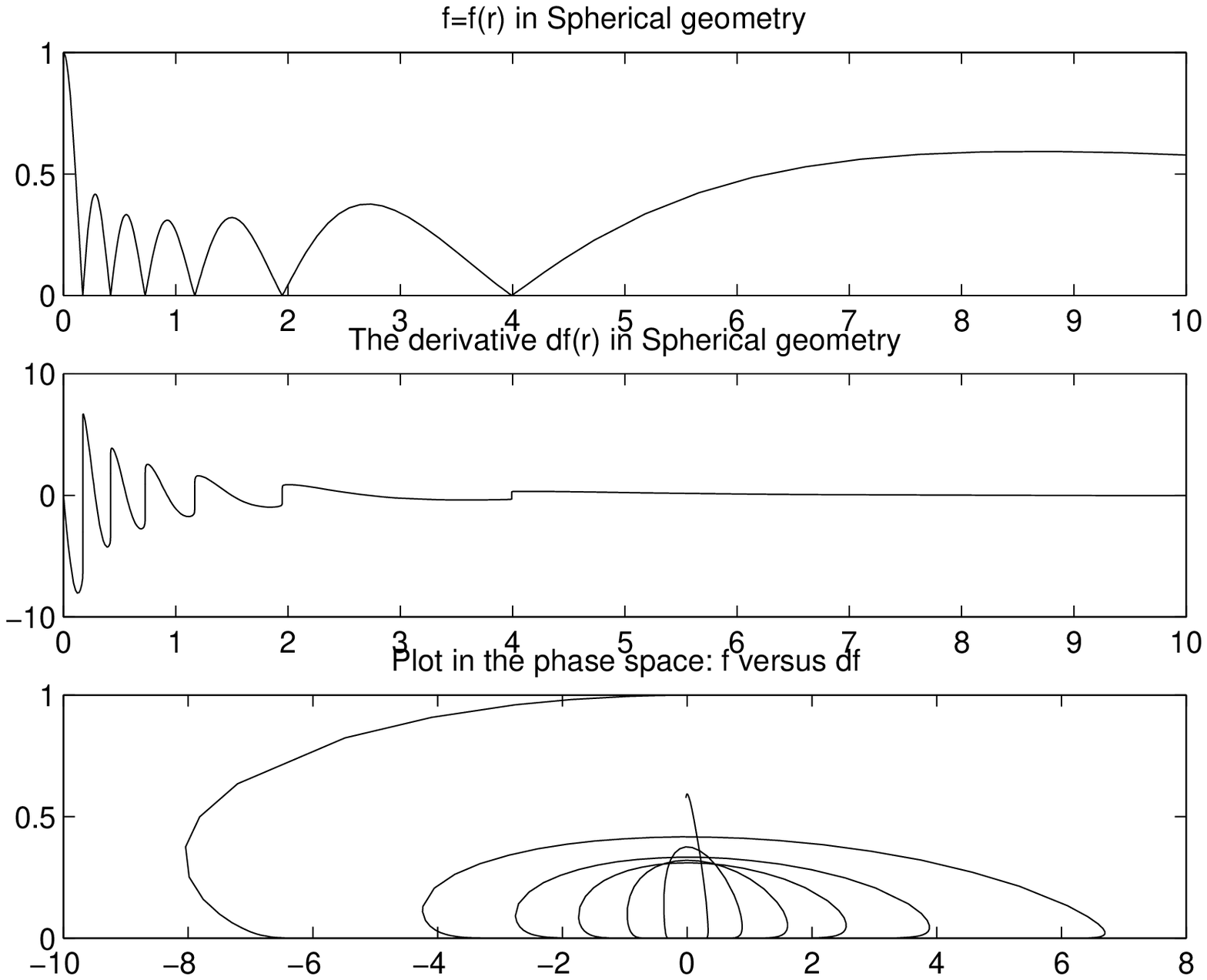}
 
\vspace{1 cm} \vbox spread 3in{}
 \includegraphics{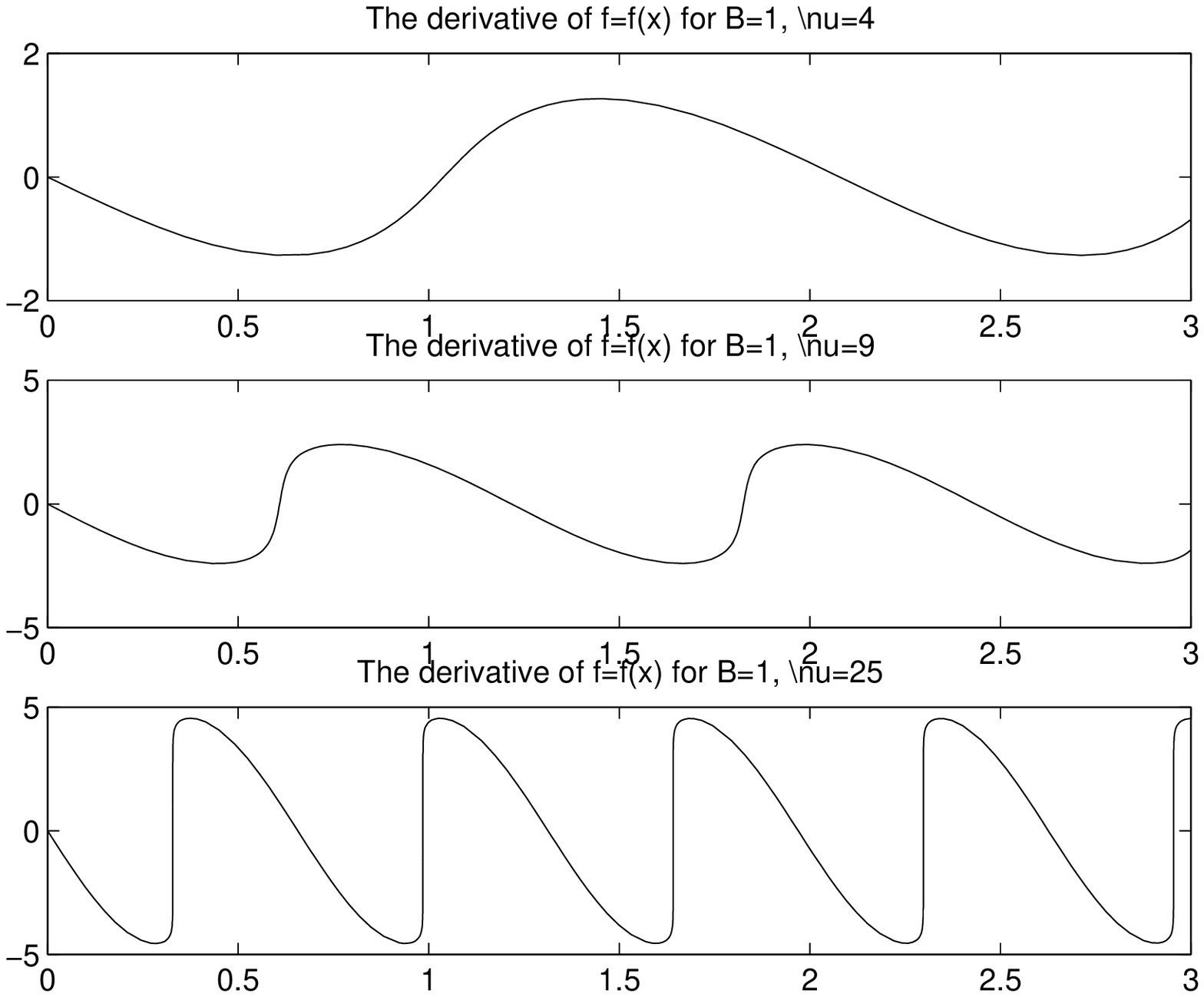} 

\vspace{1 cm} \vbox spread 3in{}
 \includegraphics{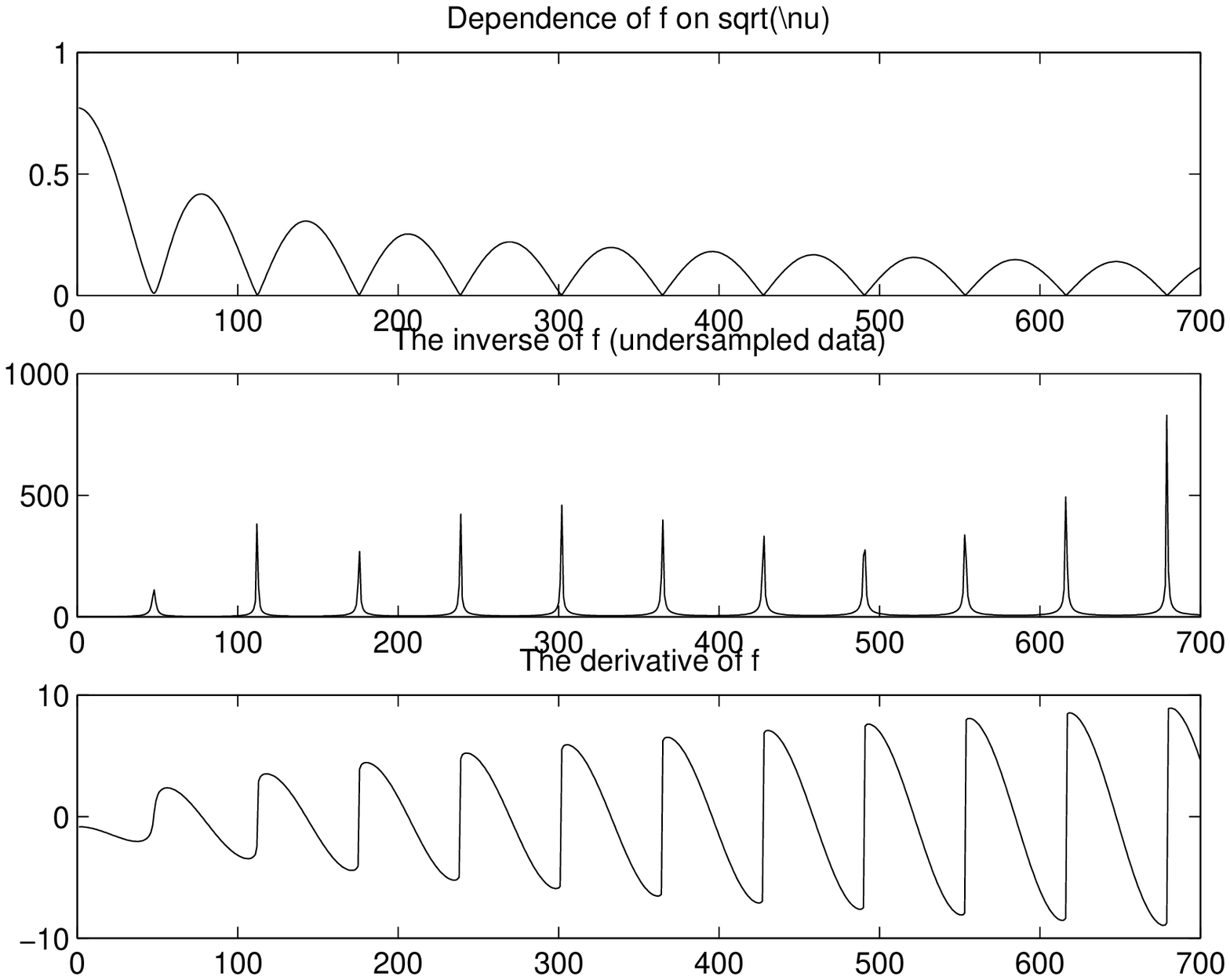}
 
%\vspace{1 cm} \vbox spread 3in{}
% \special{psfile = dfonsqrtbig.ps hoffset=36 voffset=-72 hscale=50 
%vscale=50}

\item  In conclusion, we want to shed some light on simple features
of solutions of the system, considering it on a compact two-manifold
$M$. In what follows, we assume for simplicity that $Vol(M)=1$.
Now, let $f$ be a solution of the equation (\ref{twodim}).
Since $F=\star\frac{B}{f}$ is
the curvature of a connection on a principal $U(1)$-bundle, we know that
the first Chern class (the topological quantum number) $\int F = N\in
2\pi Z$, and therefore
\begin{equation}
\int\frac{B}{f} = \int F = N \in 2\pi Z.
\label{cond}
\end{equation}
Let us also note that if $f$ satisfies the equation (\ref{twodim})
and the condition (\ref{cond}), then so does each of its  multiples 
(with a different value of $B$).
Thus, we may assume without loss of generality that $\int\frac{1}{f}=1$, and
the original equation becomes
\begin{equation}
 -\bigtriangleup f +\frac{N^2}{f}=\nu f.
\label{quant}
\end{equation}
Integrating both sides of the equation, we obtain
$N^2\int\frac{1}{f}=\nu\int f$, and since $\int\frac{1}{f}=1, Vol(M)=1$, we
have by convexity of the function $x\longrightarrow\frac{1}{x}$ that 
$\frac{N^2}{\nu} =\int f\geq\frac{1}{\int\frac{1}{f}}=1$, and so
\[ \nu\leq N^2.
\]
Multiplying (\ref{twodim}) by $f$ and integrating both sides again,
we obtain
\[\int |df|^2  + N^2 = \nu \int f^2.
\]
On the other hand $ \int f^2 \geq (\int f)^2 = \frac{N^4}{\nu ^2}$, and
combining this with the equation above, we finally obtain
\begin{tw} For solutions of (\ref{twodim})
with the topological quantum number $N$, and such that
$\int\frac{1}{f}=1$, we have
\[
  \int |df|^2 \geq N^2 (\frac{N^2}{\nu}-1) \geq 0,
\]
where both inequalities become identities if and only if $f$ is the
constant solution.
\end{tw}
Therefore one expects that the solutions for a fixed bundle become more
``convolved" as the secondary quantum number $\nu$ gets smaller.
It remains an open problem whether the set of eigenvalues $\nu$ is 
discrete
on compact manifolds.

\end{enumerate}


\begin{thebibliography}{9}
\bibitem{pr} R. E. Prange, S. M. Girvin, Eds., The Quantum Hall Effect,
Springer-Verlag, 1990
\bibitem{sch} D. Schoenberg, Magnetic Oscillations in Metals, Cambridge
University Press, 1984
\bibitem{sow1} A. Sowa, Abstracts of the AMS, Vol. 15, No. 3, p. 395, 1994 
\bibitem{sow2} A. Sowa, Geometric Unification of Yang-Mills and 
Shr\"{o}dinger Equations, thesis, The City University of New York
\bibitem{sow3} A. Sowa, On an Equation Arising From the Geometry of 
Riemannian Submersions, preprint
 \end{thebibliography}
\end{document}